\documentclass[12pt]{article}

\usepackage{amsmath,amssymb,amsfonts,graphics,graphicx,amscd,amsfonts,epsfig,color}
\usepackage{subfig}
\usepackage{here}
\usepackage{cite}
\newcommand {\beq} {\begin{equation}}
\newcommand {\eeq} {\end{equation}}
\newcommand {\beqa}{\begin{eqnarray}}
\newcommand {\eeqa}{\end{eqnarray}}
\newcommand {\nn} {\nonumber}

\newcommand{\bbZ}{{\mathbb Z}}

\makeatletter
    
    \@addtoreset{equation}{section}
  \makeatother

\allowdisplaybreaks[4]

\date{}
\begin{document}

\begin{flushright} 
KEK-TH-1981
\end{flushright} 

\vspace{1.0cm}

\begin{center}
{\bf \Large A new method for probing the late-time dynamics
in the Lorentzian type IIB matrix model}
\end{center}

\vspace{1.5cm}

\begin{center}

         Takehiro A{\sc zuma}$^{a}$\footnote
          {
 E-mail address : azuma@mpg.setsunan.ac.jp},  
         Yuta I{\sc to}$^{b}$\footnote
          {
 E-mail address : yito@post.kek.jp},  
         Jun N{\sc ishimura}$^{bc}$\footnote
          {
 E-mail address : jnishi@post.kek.jp} 
and
         Asato T{\sc suchiya}$^{d}$\footnote
          {
 E-mail address : tsuchiya.asato@shizuoka.ac.jp} 

\vspace{0.5cm}

$^a${\it Institute for Fundamental Sciences, Setsunan University, \\
17-8 Ikeda Nakamachi, Neyagawa, Osaka, 572-8508, Japan}

$^b${\it KEK Theory Center, 
High Energy Accelerator Research Organization,\\
1-1 Oho, Tsukuba, Ibaraki 305-0801, Japan} 

$^c${\it Graduate University for Advanced Studies (SOKENDAI),\\
1-1 Oho, Tsukuba, Ibaraki 305-0801, Japan} 

$^d${\it Department of Physics, Shizuoka University,\\
836 Ohya, Suruga-ku, Shizuoka 422-8529, Japan}

\end{center}

\vspace{1.5cm}

\begin{center}
  {\bf abstract}
\end{center}

\noindent The type IIB matrix model has been investigated
as a possible nonperturbative formulation of superstring theory.
In particular, it was found 
by Monte Carlo simulation of the Lorentzian version that
the 9-dimensional rotational symmetry of the spatial matrices
is broken spontaneously to the 3-dimensional one
after some ``critical time''.
In this paper we develop a new simulation method
based on the effective theory for
the submatrices corresponding to the late time.
Using this method, one can obtain the results
for $N\times N$ matrices
by simulating matrices typically of the size $O(\sqrt{N})$.
%
%
%
%
We confirm the validity of this method and
demonstrate its usefulness in simplified models.
%

\newpage

\section{Introduction}

The type IIB matrix model was proposed as a possible
nonperturbative formulation of superstring theory 
in 1996 \cite{Ishibashi:1996xs}.
As such, it has the potential to explain the beginning of our Universe
and the emergence of the real world as we observe it today.
Among many works in this direction, Ref.~\cite{Kim:2011cr}
investigated the Lorentzian version of the model by Monte Carlo simulation,
and demonstrated
that (3+1)-dimensional expanding Universe naturally appears from the model.
This implies, in particular, that the 9-dimensional rotational symmetry
of the model is broken spontaneously to the 3-dimensional one 
at certain time.   
(See Refs.~\cite{Kim:2011ts,Kim:2012mw,Ito:2013ywa,Ito:2015mxa,Ito:2017rcr,%
Nishimura:2011xy,Anagnostopoulos:2013xga,Anagnostopoulos:2015gua,%
Chaney:2015ktw,Chaney:2015mfa,Stern:2014aqa,%
Steinacker:2016vgf,Steinacker:2010rh,Klammer:2009ku,%
Yang:2015vna,Yang:2008fb,Tomita:2015let,Chatzistavrakidis:2014vsa}
for related works.)
It was also suggested from Monte Carlo simulation of simplified models
that the expansion is exponential at the beginning and it turns into a
power law at late times \cite{Ito:2013ywa,Ito:2015mxa}.
In order to investigate such late-time behaviors,
one needs to perform Monte Carlo simulation with large matrices,
which is prohibitively time-consuming for the original model
mainly due to the existence of fermionic matrices.

In this paper we develop a new simulation method for probing 
the late-time dynamics in the Lorentzian type IIB matrix model
with relatively small matrices.
This is made possible by simulating the effective theory
for the submatrices corresponding to the late time.
Typically the size of the submatrices corresponding to the time
after the SSB of the SO(9) symmetry
is $O(\sqrt{N})$, where $N$ is the size of the whole matrices.
Since the computational cost of the original model is roughly
$O(N^5)$, the new method, if applied to these submatrices,
enables us to extract essentially the same information
with the cost of $O(N^{5/2})$, which is slightly less than
the $O(N^{3})$ cost required for simulating the bosonic counterparts.

Our method is based on the idea that the effective theory for the 
submatrices has a simple action dictated by the SU($N$) symmetry
and the SO(9) rotational symmetry \cite{Ito:2013ywa}.
An important new ingredient
here
is that we can actually make full use of the SO(9,1) Lorentz symmetry
instead of just the SO(9) rotational symmetry.
This is important, in particular, in applying the method to
the models with fermionic matrices since otherwise we have to 
fine-tune one parameter in the fermionic action, which controls
the ratio of the temporal part and the spatial part.
In fact, the SO(9,1) Lorentz symmetry is softly
broken by the infrared cutoffs, which are inevitably introduced
to make the model well-defined, and it is not obvious
that we can impose the SO(9,1) Lorentz symmetry on the effective
theory. Our results indicate that this is indeed the case.

The rest of this paper is organized as follows.
In section \ref{sec:review} we briefly review 
the type IIB matrix model.
In section \ref{sec:eff-theory}
we describe the new method, which is based
on the effective theory for the submatrices
corresponding to the late time.
In section \ref{sec:The-application-of} we demonstrate the usefulness
of the method by applying it to a bosonic model.
In section \ref{sec:Properties-of-the} we discuss how
one can tune the parameters in the effective theory
to optimize the efficiency of the method.
Section \ref{sec:summary} is devoted to a summary and discussions.
In appendix \ref{sec:appendix_details}
we give the detailed information used to make the plots in this paper.
In appendix \ref{sec:appendix_6dsusy} we present the results of the 
analysis in section \ref{sec:eff-theory} 
for a supersymmetric model including fermionic matrices.

\section{Brief review of the type 
IIB matrix model\label{sec:review}}

The type IIB matrix model \cite{Ishibashi:1996xs}
is composed of
10 bosonic $N\times N$ matrices 
$A_{\mu}$ $\left(\mu=0,\ldots,9\right)$ 
and 16 fermionic $N\times N$ matrices 
$\Psi_{\alpha}$ $\left(\alpha=1,\ldots,16\right)$, 
which are both traceless Hermitian.
Its action is given by
\begin{eqnarray}
S & = & S_{{\rm b}}+S_{{\rm f}} \ ,
\label{eq:S_likkt}\\
S_{{\rm b}} & = & 
\frac{1}{4g^{2}}{\rm Tr}
\left(\left[A_{\mu},A_{\nu}\right]
\left[A^{\mu},A^{\nu}\right]\right) \ ,
\label{eq:Sb}\\
S_{{\rm f}} & = & 
-\frac{1}{2g^{2}}{\rm Tr}
\left(\Psi_{\alpha}\left(\mathcal{C}
\Gamma^{\mu}\right)_{\alpha\beta}
\left[A_{\mu},\Psi_{\beta}\right]\right) \ ,
\label{eq:Sf-1}
\end{eqnarray}
where 
the indices $\mu,\nu$
are contracted using the Lorentzian metric 
$\eta_{\mu\nu}={\rm diag}\left(-1,1,\ldots,1\right)$.
We have also introduced
the 10d gamma-matrices $\Gamma^{\mu}$
after the Weyl projection and the charge conjugation
matrix $\mathcal{C}$.
The parameter $g$ is merely a scale parameter
in this model 
instead of being a coupling constant
since it can be absorbed by rescaling $A_{\mu}$ and
$\Psi$ appropriately. 
The Euclidean version (studied e.g., in 
Refs.~\cite{Nishimura:2011xy,Anagnostopoulos:2013xga})
can be obtained by making the 
``Wick rotation'' $A_0 = i A_{10}$, where $A_{10}$ 
is supposed to be Hermitian.

The partition function 
for the Lorentzian version
is proposed in Ref.~\cite{Kim:2011cr} as 
\begin{equation}
Z=\int dAd\Psi\, e^{iS}
\label{Z-Likkt1}
\end{equation}
with the action \eqref{eq:S_likkt}. 
The ``$i$'' in front of the action
is motivated from the fact that
the string world-sheet metric should also have 
a Lorentzian signature. 
By integrating out the fermionic matrices,
we obtain the Pfaffian 
\begin{equation}
\int d\Psi\, e^{iS_{{\rm f}}} =
{\rm Pf}\mathcal{M}\left(A\right)  \ ,
\end{equation}
which is real unlike in the Euclidean case \cite{Anagnostopoulos:2013xga}.
Note also that the bosonic action \eqref{eq:Sb}
can be written as 
\begin{eqnarray}
S_{{\rm b}}  =  
\frac{1}{4g^{2}}{\rm Tr}\left(F_{\mu\nu}F^{\mu\nu}\right)
 =  \frac{1}{4g^{2}}
\left\{ {\rm -2Tr}\left(F_{0i}\right)^{2}+
{\rm Tr}\left(F_{ij}\right)^{2}\right\} \ ,
\label{decomp-Sb}
\end{eqnarray}
where we have introduced
the Hermitian matrices $F_{\mu\nu}=i\left[A_{\mu},A_{\nu}\right]$.
Since the two terms
in the last expression
have opposite signs,
$S_{{\rm b}}$ is not positive semi-definite,
and it is not bounded from below.

In order to make the partition function \eqref{Z-Likkt1} finite,
one needs to introduce infrared cutoffs 
in both the temporal and spatial directions, 
for instance, as \cite{Kim:2011cr}
\begin{eqnarray}
\frac{1}{N}{\rm Tr}\left(A_{0}\right)^{2} 
& \leq & \kappa\frac{1}{N}{\rm Tr}
\left(A_{i}\right)^{2} \ ,\label{eq:t_cutoff}\\
\frac{1}{N}{\rm Tr}\left(A_{i}\right)^{2} 
& \leq & \Lambda^{2} \ .
\label{eq:s_cutoff}
\end{eqnarray}
Recently we have found it important to
generalize this form of the infrared cutoffs slightly 
in order to achieve universality in the large-$N$ 
limit \cite{Ito:2017rcr}.
However, in the present methodological paper, we use the 
above form for simplicity.

After some manipulation and rescaling of $A_{\mu}$, 
we can rewrite the partition function (\ref{Z-Likkt1})
as \cite{Kim:2011cr} (See 
appendix A of Ref.~\cite{Ito:2013ywa} for a refined argument.)
\begin{alignat}{3}
Z=\int dA\,{\rm Pf}\mathcal{M}(A) \delta\left(
\frac{1}{N}{\rm Tr}\left(F_{\mu\nu}F^{\mu\nu}\right)\right)
\delta\left(\frac{1}{N}{\rm Tr}\left(A_{i}\right)^{2}-L^2 \right)
\theta\left(\kappa L^2 -\frac{1}{N}{\rm Tr}\left(A_{0}\right)^{2}\right)\ ,
\label{Z-Likkt3}
\end{alignat}
where $\theta\left(x\right)$ is the Heaviside step function.
The scale parameter $L$ can be set to unity without loss of generality.
This form allows us to perform Monte Carlo simulation
of the Lorentzian model without the sign problem 
unlike the Euclidean model.\footnote{Strictly speaking,
the Pfaffian ${\rm Pf}\mathcal{M}$ 
in (\ref{Z-Likkt3})
can change its sign,
but it is found that configurations with positive Pfaffian
dominate at large $N$.\label{footnote:phase-quench}}
See appendix B of Ref.~\cite{Ito:2013ywa} for the details of
Monte Carlo simulation.

It turns out \cite{Kim:2011cr} that
one can extract a time-evolution
from configurations generated by simulating (\ref{Z-Likkt3}).
For that purpose, we first use
the ${\rm SU}\left(N\right)$ symmetry of the model
to bring the temporal matrix $A_{0}$ into the diagonal form
\begin{equation}
A_{0}={\rm diag}\left(\alpha_{1},\ldots,\alpha_{N}\right)\ ,
\quad \quad
{\rm where~} \alpha_{1}<\cdots<\alpha_{N} \ .
\label{eq:diagonal gauge}
\end{equation}
In this basis,
the spatial matrices $A_{i}$ are found to have
a band-diagonal structure.
More precisely, there exists some integer $n$ such that
the elements of spatial matrices
$\left(A_{i}\right)_{ab}$ for $\left|a-b\right|>n$ are 
much smaller than those for $\left|a-b\right|\leq n$.
Based on this observation,
we may naturally consider $n\times n$ matrices
\begin{equation}
(\bar{A}_{i})_{IJ}(t)
\equiv (A_{i})_{K+I,K+J} \ ,
\label{eq:def_abar}
\end{equation}
as representing the state of the universe at some time $t$, where
$I,J=1,\ldots , n$ and $K=0,1,\ldots , N-n$.
The time $t$ 
is defined by 
\begin{equation}
t=\frac{1}{n}\sum_{I=1}^{n}\alpha_{K+I}
\label{eq:def_t}
\end{equation}
corresponding to the $n\times n$ matrices $\bar{A}_{i}$. 
For example,
we can define the extent of space at time $t$ as 
\begin{equation}
R^{2}(t)= \left\langle \frac{1}{n}{\rm tr}\sum_{i}
\left(\bar{A}_{i}\left(t\right)\right)^{2}\right\rangle \ ,
\label{eq:def_rsq}
\end{equation}
where the symbol ${\rm tr}$ represents
a trace over the $n\times n$ block.
We also define
the ``moment of inertia tensor'' 
\begin{equation}
T_{ij}(t)
=\frac{1}{n}{\rm tr}
\left(\bar{A}_{i}\left(t\right)\bar{A}_{j}\left(t\right)\right) \ ,
\label{eq:def_tij}
\end{equation}
which is a $9\times9$ real symmetric matrix. 
The eigenvalues of $T_{ij}\left(t\right)$,
which we denote by $\lambda_{i}\left(t\right)$ with the ordering
\begin{equation}
\lambda_{1}\left(t\right)>\lambda_{2}
\left(t\right)>\cdots>\lambda_{9}\left(t\right) \ ,
\end{equation}
represent the spatial extent in each of 
the nine directions at time $t$.
Note that the expectation values 
$\left\langle \lambda_{i}\left(t\right)\right\rangle $
tend to be equal in the large-$N$ limit if the SO(9) symmetry is
not spontaneously broken. 
This is the case at early times of the time-evolution.
After a critical time $t_{{\rm c}}$, however,
it was found \cite{Kim:2011cr} that
three largest eigenvalues
$\left\langle \lambda_{i}\left(t\right)\right\rangle$ 
($i=1$, $2$, $3$)
become significantly larger than the others,
which implies that
the SO(9) symmetry is spontaneously broken down to SO(3).

\begin{figure}[t]
\centering{}
\includegraphics{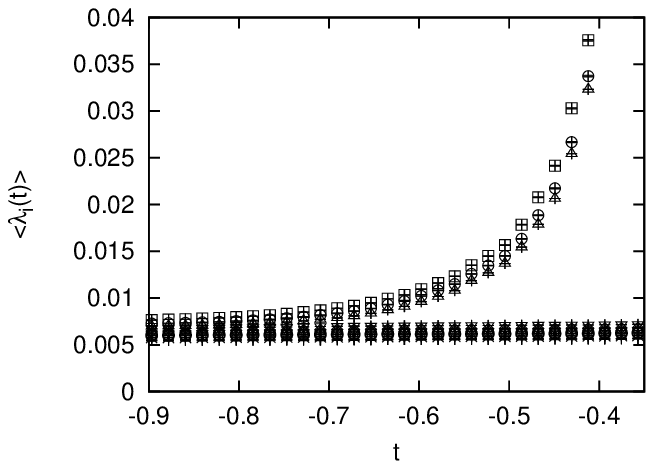}
\includegraphics{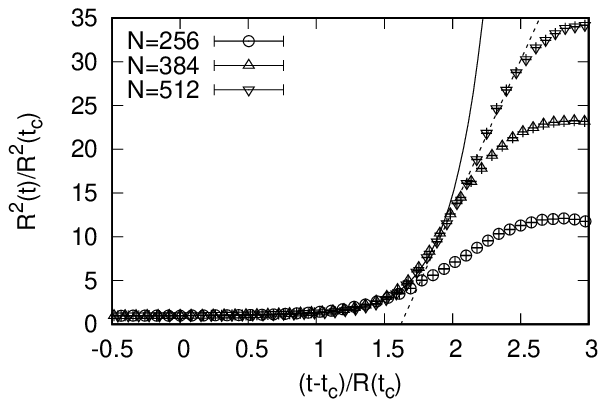}
\caption{
(Left) The expectation values 
$\langle \lambda_{i}\left(t\right) \rangle$
of the nine eigenvalues of $T_{ij}\left(t\right)$ 
are plotted against $t$ for the 10d bosonic model with $N=512$.
(Right) The extent of space 
$R^{2}\left(t\right)/R^2\left(t_{\rm c}\right)$
is plotted against 
$x = \left(t-t_{\rm c}\right)/R\left(t_{\rm c}\right)$ 
for the 10d bosonic model with 
$N=256$, $384$ and $512$. 
%
%
The solid line is a fit of the $N=512$ data to 
$R^{2}\left(t\right)/R^{2}\left(t_{\rm{c}}\right)=
a+\left(1-a\right)\exp\left(bx\right)$ 
for $ 1.0 \le x \le 1.85$,
which gives $a=0.9957(5)$ and $b=4.03(7)$. 
The dashed line is a fit of the $N=512$ data
to $R^{2}\left(t\right)/R^{2}\left(t_{\rm{c}}\right)=cx+d$ 
for $1.85 \le x \le 2.5$,
which gives $c=34.3(6)$ and $d=-55(1)$.
\label{fig:10d-bosonic-original}
}
\end{figure}

\begin{figure}[t]
\centering{}
\includegraphics[width=10cm]{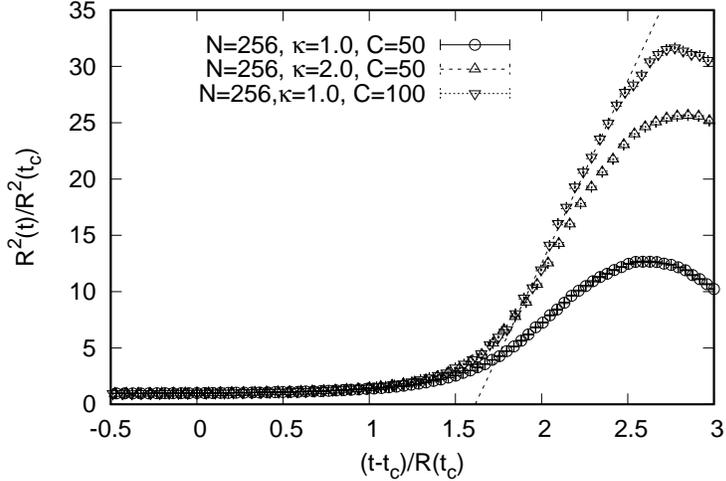}
\caption{
The extent of space 
$R^{2}\left(t\right)/R^{2}\left(t_{\mathrm{c}}\right)$
are plotted against 
$x=\left(t-t_{\mathrm{c}}\right)/R\left(t_{\mathrm{c}}\right)$
for the effective theory of the 10d bosonic model 
with $N=256$ and various values of $\kappa$ and $C$.
The dashed line represents a fit of the data points 
for $(\kappa, C)=(1.0,100)$
to $R^{2}\left(t\right)/R^{2}\left(t_{\mathrm{c}}\right)=cx+d$ 
with $1.85\leq x\leq2.5$, which gives $c=32.4(6)$ and $d=-52(1)$.
\label{fig:10d-bosonic}
}
\end{figure}

The block size $n$ used in calculating
quantities such as (\ref{eq:def_rsq}) and (\ref{eq:def_tij})
by Monte Carlo simulation
is determined as described in section 5 of Ref.~\cite{Ito:2015mxa}.
In appendix \ref{sec:appendix_details}
we present the block size used in making the plots in 
Figs.~\ref{fig:10d-bosonic-original},
\ref{fig:10d-bosonic}, 
\ref{fig:c-dep_and_k-dep} and \ref{fig:eps_and_delta}.
There, we also present the values obtained
for the critical time $t_{{\rm c}}$ and the corresponding 
extent of space $R(t_{\rm c})$, which are also needed 
in making these plots.

In Fig.~\ref{fig:10d-bosonic-original} we show the results \cite{Ito:2015mxa}
for the bosonic model, which is obtained by omitting the 
Pfaffian in (\ref{Z-Likkt3}).
As we find in Ref.~\cite{Ito:2015mxa},
the bosonic model is well defined without
the infrared cutoff (\ref{eq:t_cutoff})
for the temporal matrix $A_0$.
Therefore, the constraint
$\theta\left(\kappa -\frac{1}{N}{\rm Tr}\left(A_{0}\right)^{2}\right)$
for the temporal matrix $A_0$
in (\ref{Z-Likkt3}) can be omitted.
On the left, we plot the expectation values
$\langle \lambda_{i}\left(t\right) \rangle$
against $t$ for $N=512$.
We observe the 
SSB
from SO(9) to SO(3) at a critical time $t_{{\rm c}}$
similarly to the original Lorentzian type IIB matrix 
model.
On the right, 
the extent of space $R^{2}\left(t\right)/R^{2}\left(t_{\mathrm{c}}\right)$
is plotted 
against $(t-t_{\mathrm{c}})/R(t_{\mathrm{c}})$
for various $N \le 512$.
We observe a clear scaling behavior.
It is found that 
the behavior of $R^2(t)$ at $t > t_{\rm c}$
can be fitted to an exponential function
only for a finite range, and it seems to slow down
at later times in contrast to the exponential behavior
observed in Ref.~\cite{Ito:2013ywa}
in a simplified model for early times.
We can actually fit our $N=512$ data
within $1.85 \le (t-t_{\rm c})/R(t_{\rm c}) \le 2.5$
to a straight line,
which corresponds to the power-law expansion
\begin{equation}
R\left(t\right)\propto t^{1/2} 
\label{power-law-exp}
\end{equation}
reminiscent 
of the expanding behavior of the Friedmann-Robertson-Walker universe
in the radiation dominated era.
This is interesting given that 
the bosonic model used to obtain these results is expected
to capture the late-time dynamics of the original model \cite{Ito:2015mxa}.

In Monte Carlo simulation, 
we find that the $\bbZ _2$ symmetry $A_0 \mapsto -A_0$ is not broken
spontaneously, and hence the time-evolution we obtain 
from the generated matrices
respects the time-reversal symmetry. 
This by no means implies that the Universe is doomed to end with
a Big Crunch because one has to take the large-$N$ limit.
In that limit, it could be that 
the turning point $t=0$ is infinitely away, in physical units,
from the ``critical time'' $t_{\rm c}$, at which the SSB occurs; 
namely $|t_{\rm c}|/R(t_{\rm c}) \rightarrow \infty$ as $N\rightarrow \infty$.
In other words, the results for finite $N$ allow us to probe only
the early-time dynamics of the model, 
and the appearance of the turning point at $t=0$ can be 
merely a finite-$N$ artifact.
In order to probe the late-time dynamics, we need to increase $N$,
which is hard in the original model because 
the computational cost\footnote{In order to make one trajectory
in the Hybrid Monte Carlo algorithm,
the original model requires $O(N^5)$ arithmetic operations.
The reason for this is that the number of iterations required for
the convergence of the conjugate gradient method
used to implement the effects of fermions grows as $O(N^2)$.
Since matrix multiplication requires 
$O(N^3)$ arithmetic operations, we get $O(N^5)$.}
increases as $O(N^5)$.

It should be noted at this point that
the number of data points in the SSB region 
in Fig.~\ref{fig:10d-bosonic-original} (Right)
is 61, 73, 83 for $N=256$, $384$, $512$, respectively,
which roughly grows as $4 \sqrt{N}$.
This hints at a possible method for simulating
the SSB region more efficiently.

\section{The effective theory for the submatrices}
\label{sec:eff-theory}

\begin{figure}[t]
\centering{}
\includegraphics{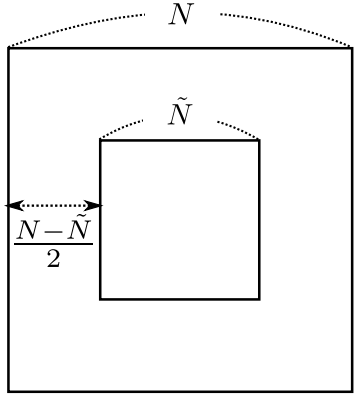}
\caption{
This figure shows how we cut out
the $\tilde{N} \times \tilde{N}$ submatrices
from the $N \times N$ matrices
that appear in the Lorentzian type IIB matrix model.
\label{fig:image}
}
\end{figure}

In this section we consider the effective theory for 
the submatrices corresponding to the late times.
In view of the discussion given in the previous section,
we choose the SU($N$) basis in which the temporal matrix
$A_0$ takes the diagonal form (\ref{eq:diagonal gauge}),
and cut the $\tilde{N} \times \tilde{N}$ submatrices
$\tilde{A}_{\mu}$ 
out of the whole matrices $A_{\mu}$
as depicted in Fig.~\ref{fig:image}.
More explicitly, we define the submatrices $\tilde{A}_{\mu}$ as
\begin{equation}
(\tilde{A}_{\mu})_{ab}
=\left(A_{\mu}\right)_{s+a,s+b}\,,\quad s\equiv\frac{N-\tilde{N}}{2}\ ,
\end{equation}
where $a,b=1 , \ldots , \tilde{N}$.
Using these submatrices, we define the expectation values
\begin{equation}
\tilde{C}\equiv
\left\langle 
\frac{1}{\tilde{N}}\mathrm{Tr}
(\tilde{F}_{\mu\nu}\tilde{F}^{\mu\nu})\right\rangle \ ,\quad
\tilde{L}^{2}\equiv
\left\langle \frac{1}{\tilde{N}}\mathrm{Tr}
(\tilde{A}_{i})^{2}\right\rangle \ ,\quad
\tilde{\kappa}\tilde{L}^{2}\equiv
\left\langle \frac{1}{\tilde{N}}\mathrm{Tr}
(\tilde{A}_{0} )^{2}\right\rangle
\label{eq:observables}
\end{equation}
with the matrices $A_{\mu}$ obtained by simulating the original matrix model,
where we have defined
$\tilde{F}_{\mu\nu}=i[\tilde{A}_{\mu},\tilde{A}_{\nu}]$.

Let us then consider the effective theory for
the $\tilde{N}\times\tilde{N}$ submatrices $\tilde{A}_{\mu}$.
Since the expectation values (\ref{eq:observables})
should be reproduced from the effective theory,
a natural candidate for the effective theory is
\begin{alignat}{3}
 & Z_{\mathrm{eff}} [\tilde{N};N,\kappa ]  \nn \\
&= \int d\tilde{A}\,\mathrm{Pf}\mathcal{M} (\tilde{A})
\delta\left(\frac{1}{\tilde{N}}
\mathrm{Tr}
(\tilde{F}_{\mu\nu}\tilde{F}^{\mu\nu})-\tilde{C}\right)
\delta\left(\frac{1}{\tilde{N}}
\mathrm{Tr}
(\tilde{A}_{i})^{2}
-\tilde{L}^{2}\right)
\theta\left(\tilde{\kappa}\tilde{L}^{2}-\frac{1}{\tilde{N}}
\mathrm{Tr}(\tilde{A}_{0})^{2}\right) \  .
\label{eq:effective_theory}
\end{alignat}
The only difference
from the original Lorentzian 
type IIB matrix model (\ref{Z-Likkt3})
is the appearance of the parameter $\tilde{C}$.
In order to verify this ansatz,
we calculate the 
quantities such as (\ref{eq:def_rsq}) and (\ref{eq:def_tij})
using the effective theory and see whether the results agree with the ones
obtained from the original model.

\begin{figure}[t]
\centering{}
\includegraphics{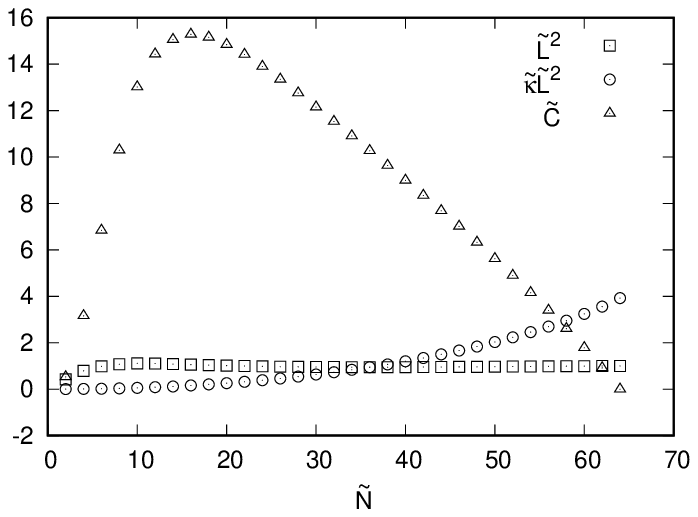}\includegraphics{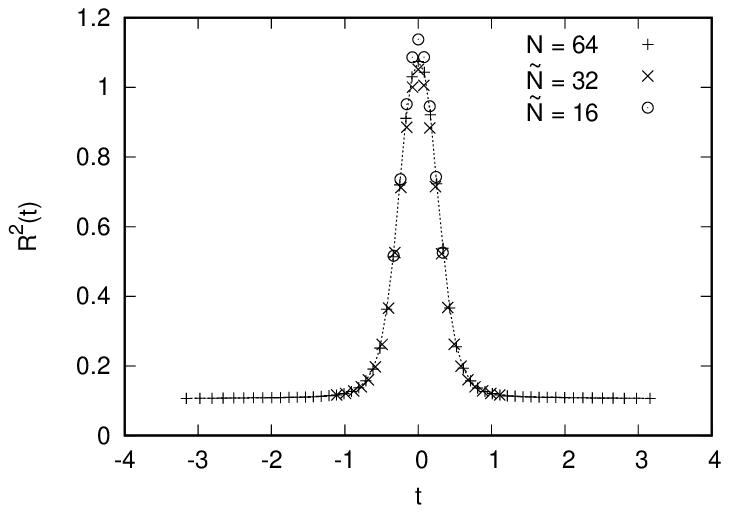}
\caption{
(Left) The expectation values $\tilde{L}^{2}$, $\tilde{\kappa}\tilde{L}^{2}$
and $\tilde{C}$ defined in (\ref{eq:observables}) are plotted against
the size $\tilde{N}$ of the submatrices
for the 6d version of the bosonic model with $N=64$. 
(Right) The extent of space $R^{2}(t)$
is plotted against $t$ for the 6d version of the bosonic
model. The dotted line, which is drawn to guide the eye,
connects the data points (plus sign) for the original model with $N=64$.
The circles and crosses represent the results of 
$\tilde{R}^{2} (\tilde{t})$ for
the effective theory with $\tilde{N}=16$ and $32$, respectively.
The block size used in making this plot is $n=8$ for all cases. 
\label{fig:test_rgm}
}
\end{figure}

For simplicity, here we consider 
the 6d version of the bosonic model,\footnote{Let us recall
that in the bosonic models, there is no need to introduce 
the infrared cutoff (\ref{eq:t_cutoff})
for the temporal matrix $A_0$.
However, in the corresponding
effective theory like (\ref{eq:effective_theory}),
we need to introduce 
$\theta\left(\tilde{\kappa}\tilde{L}^{2}-\frac{1}{\tilde{N}}
\mathrm{Tr}(\tilde{A}_{0})^{2}\right)$,
which represents 
the infrared cutoff for the temporal matrix $A_0$.
Also, when the bosonic models are considered
in the argument for justifying the use of the 
effective theory (\ref{eq:rescaled_zeff}) without simulating
the original model, we assume that the parameter $\kappa$
is introduced also in the original model,
where $\kappa$ should be smaller than the value of 
$\frac{1}{N} \mathrm{Tr}(A_{0})^{2}$ that is otherwise obtained.
}
which can be obtained by restricting ourselves to
$A_{\mu}$ $\left(\mu=0,\ldots,5\right)$.
The qualitative behavior of this simplified model 
is analogous to the original model,
and, in particular, 
the SSB from SO(5) to SO(3) occurs.
We perform simulation of this model with the matrix size $N=64$,
and plot the expectation values 
(\ref{eq:observables}) for the $\tilde{N} \times \tilde{N}$ submatrices
against $\tilde{N}$ in Fig.~\ref{fig:test_rgm} (Left).
Then, using these expectation values,
we perform simulation of the effective theory
(\ref{eq:effective_theory})
for the $\tilde{N} \times \tilde{N}$ submatrices
omitting the Pfaffian ${\rm Pf}\mathcal{M}$.
In Fig.~\ref{fig:test_rgm} (Right)
we plot the extent of space $\tilde{R}^{2}(\tilde{t})$
against $\tilde{t}$ for 
the effective theory with $\tilde{N}=16$, $32$ and 
compare it with the results for the original model with $N=64$.
We find that the effective theory indeed reproduces the
late-time behaviors of the original model correctly
except the region around $\tilde{t}=0$, which is subject to 
finite-$N$ effects anyway.
Note, in particular, that all the data points for $\tilde{N}=16$
lie in the region where the SSB of SO(5) occurs.
In appendix \ref{sec:appendix_6dsusy}
we present the results of the same analysis
for the 6d version of the supersymmetric model,
which includes fermionic matrices.

So far, we have determined the values of $\tilde{C}$,
$\tilde{L}$ and $\tilde{\kappa}$ to be used
in the effective theory (\ref{eq:effective_theory})
for the $\tilde{N}\times \tilde{N}$ submatrices 
by simulating the original model with the matrix size $N$, which 
is larger than $\tilde{N}$.
However, the real virtue of this method lies in the fact that
it can be used without simulating the original model.
In order to see that, let us first consider a theory
\begin{alignat}{3}
\hat{Z}
= \int d\hat{A} \,\mathrm{Pf}\mathcal{M}(\hat{A})
\delta\left(\frac{1}{\hat{N}}
\mathrm{Tr}
(\hat{F}_{\mu\nu}\hat{F}^{\mu\nu})-\hat{C}\right)
\delta\left(\frac{1}{\hat{N}}
\mathrm{Tr} (\hat{A}_{i} )^{2}-1\right)
\theta\left(\hat{\kappa}-\frac{1}{\hat{N}}
\mathrm{Tr}(\hat{A}_{0})^{2}\right) \ ,
\label{eq:rescaled_zeff}
\end{alignat}
which generalizes\footnote{It is an interesting historical remark
that the same type of generalization has been proposed 
in the context of the space-time uncertainty principle 
in string theory \cite{Yoneya:1997gs}.
Note, however, that the parameter $\hat{C}$ that appears in that
context should be negative unlike in our case.}
the original theory (\ref{Z-Likkt3})
by introducing nonzero $\hat{C}$.
We have put hats on all the variables and the parameters of this theory 
to distinguish them from those in the original theory.
In order for this theory to be equivalent to
the effective theory (\ref{eq:effective_theory}),
we need $\hat{N}=\tilde{N}$,
$\hat{A}_{\mu} =\tilde{A}_{\mu}/\tilde{L}$ and
\begin{alignat}{3}
\hat{\kappa} =\tilde{\kappa} (\hat{N};N,\kappa ) \ , 
\quad \quad
\hat{C} = 
\frac{\tilde{C} (\hat{N};N,\kappa )}
{\tilde{L}^{4} (\hat{N};N,\kappa )} \ .
\label{eq:correspond}
\end{alignat}
The point here is that whatever values we choose for
$\hat{\kappa}$ and $\hat{C}$, we have two parameters
$N$ and $\kappa$ at our disposal, which make it
possible to satisfy (\ref{eq:correspond}) generically
from counting the degrees of freedom.
(Strictly speaking, this statement holds only at large $N$ since
$N$ cannot be changed continuously.)
Thus, the generalized theory (\ref{eq:rescaled_zeff}) 
can always be interpreted as
the effective theory for the submatrices of the original theory 
up to some rescaling of $A_\mu$.

\section{Demonstrating the usefulness of the method}
\label{sec:The-application-of}

In this section we demonstrate the usefulness of our method
by applying it to the 10d bosonic model. 
As we reviewed in section \ref{sec:review},
this model was studied in Ref.~\cite{Ito:2015mxa}
and the 3d space
was found to expand with the power law at late times.
We will show that this result can be reproduced by using the
effective theory with much smaller $N$ than 
that used for the original model.

Here we perform simulation of the effective theory with $N=256$.
(From now on,
we omit the hats in the effective theory (\ref{eq:rescaled_zeff}).)
In Fig.~\ref{fig:10d-bosonic} we plot the extent of space
$R^2 (t)/R^2 (t_{\mathrm{c}})$ against 
$(t-t_{\mathrm{c}})/R (t_{\mathrm{c}})$
for the effective theory\footnote{In the original 10d bosonic model,
there is no need to introduce 
the infrared cutoff (\ref{eq:t_cutoff})
for the temporal matrix $A_0$,
and we obtain 
$\left\langle \frac{1}{N}\mathrm{Tr}(A_{0})^{2}\right\rangle =4.3767(1)$
for $N=256$.
Note that the values of $\kappa$ chosen in the effective theory
are considerably smaller than this value.}
with various $\kappa$ and $C$.
The results should be compared with the results in 
Fig.~\ref{fig:10d-bosonic-original} 
for the original 10d bosonic model with $N \le 512$.
We observe a clear scaling behavior 
similarly to Fig.~\ref{fig:10d-bosonic-original}
in spite of the fixed matrix size.
We also find that the result for $\kappa=1.0$ and $C=100$
coincides with the result for the original model with $N=512$
except around the peak. 
In particular, we are able to reproduce
the power-law
expansion $R(t)/R(t_{\mathrm{c}})\sim t^{1/2}$
at late times with the matrix size much smaller
than in Fig.~\ref{fig:10d-bosonic-original}.
This demonstrates the usefulness of the new method.

\begin{figure}[t]
\centering{}
\includegraphics{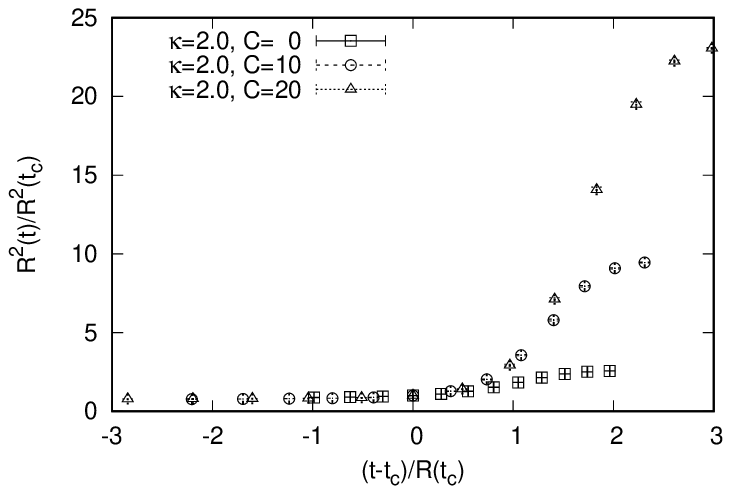}\includegraphics{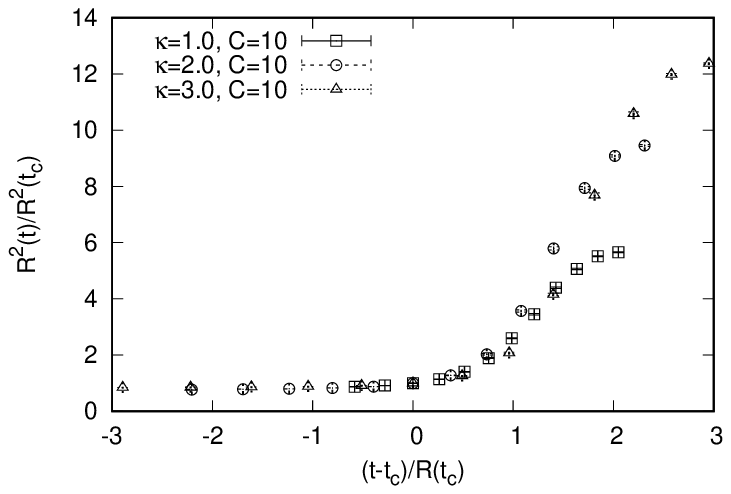}
\caption{
(Left) The extent of space 
$R^{2}\left(t\right)/R^{2}\left(t_{\mathrm{c}}\right)$
is plotted against 
$\left(t-t_{\mathrm{c}}\right)/R\left(t_{\mathrm{c}}\right)$
for the effective theory of the 6d bosonic model
with $N=32$, $\kappa=2.0$ and various $C$.
(Right) The extent of space 
$R^{2}\left(t\right)/R^{2}\left(t_{\mathrm{c}}\right)$
is plotted against 
$\left(t-t_{\mathrm{c}}\right)/R\left(t_{\mathrm{c}}\right)$
for the effective theory of the 6d bosonic model
with $N=32$, $C=10$ and various $\kappa$.
\label{fig:c-dep_and_k-dep}
}
\end{figure}

\section{How to tune the parameters in the effective theory
\label{sec:Properties-of-the}}

\begin{figure}[t]
\centering{}
\includegraphics{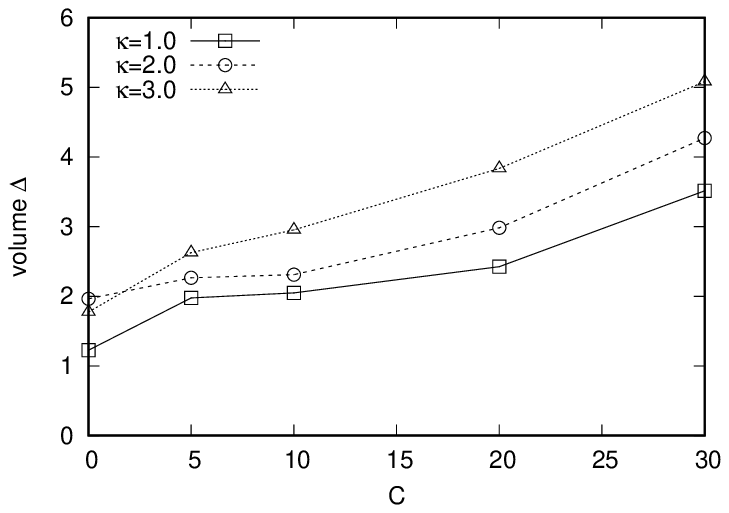}
\includegraphics{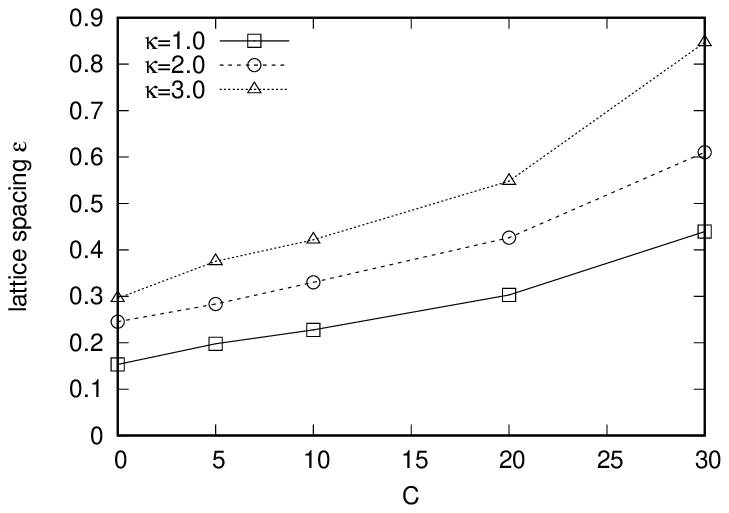}
\includegraphics{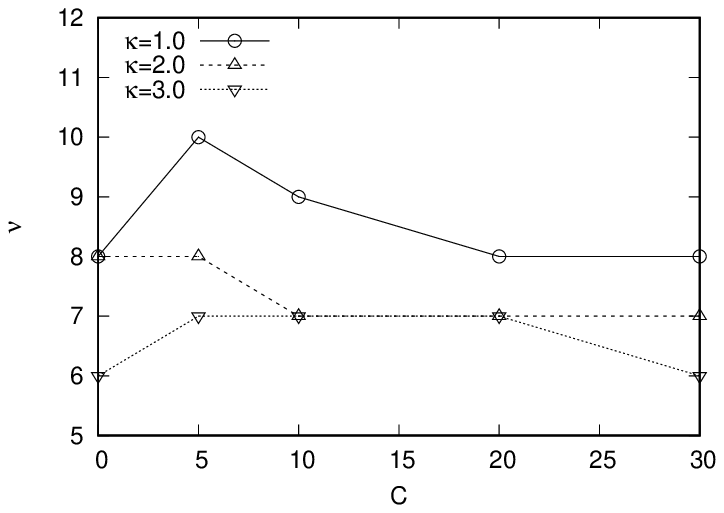}
\caption{The volume $\Delta$ (Top-Left)
and the lattice spacing $\epsilon$ (Top-Right)
are plotted against $C$ 
for the effective theory of the 6d bosonic model
with $N=32$ and various $\kappa$.
(Bottom) The number of 
data points of $R^2(t)/R^2(t_{\mathrm{c}})$
in $t_{\mathrm{c}}<t\leq t_{\mathrm{peak}}$
is plotted against $C$ 
for the effective theory of the 6d bosonic model
with $N=32$ and various $\kappa$.
\label{fig:eps_and_delta}
}
\end{figure}

We can actually optimize the efficiency of the method
by tuning the parameters in the effective theory (\ref{eq:rescaled_zeff}).
For that purpose, let us define the ``volume'' $\Delta$ 
in the temporal direction as
\begin{equation}
\Delta\equiv\frac{t_{\mathrm{peak}}-t_{c}}
{R\left(t_{\mathrm{c}}\right)} \ , 
\label{eq:def_delta}
\end{equation}
where $t_{\mathrm{peak}}$ represents the position of 
the peak\footnote{In fact, $t_{\mathrm{peak}}\simeq 0$ due to 
the time-reversal symmetry $A_0 \mapsto -A_0$.}
in $R^2(t)$.
We also define 
the ``lattice spacing'' $\epsilon$ 
in the temporal direction as
\begin{equation}
\epsilon\equiv\frac{\Delta}{\nu}\ ,
\label{eq:def_eps}
\end{equation}
where $\nu$ is the number of data points of 
$R^{2}(t)$
contained within $t_{\mathrm{c}}<t\leq t_{\mathrm{peak}}$.
This definition
represents the average horizontal spacing between the adjacent data points
of $R^2(t)$. 
Note that both $\Delta$ and $\epsilon$
can be changed by tuning $\kappa$ and $C$ in the effective theory
with the matrix size fixed.
In appendix \ref{sec:appendix_details}
we give the values of $\nu$, $\Delta$ and $\epsilon$ obtained
from the plots in Figs.~\ref{fig:10d-bosonic-original} (Right),
\ref{fig:10d-bosonic} and 
\ref{fig:c-dep_and_k-dep}.
The values plotted in Fig.~\ref{fig:eps_and_delta} are also listed.

In Fig.~\ref{fig:c-dep_and_k-dep} we plot
the extent of space 
$R^{2}\left(t\right)/R^{2}\left(t_{\mathrm{c}}\right)$ against
$x\left(t\right)\equiv
\left(t-t_{\mathrm{c}}\right)/R\left(t_{\mathrm{c}}\right)$
for various $C$ with fixed $\kappa$ (Left)
and for various $\kappa$ with fixed $C$ (Right)
in the effective theory of the 6d bosonic model with $N=32$.
Here we plot only the data points for $t\leq t_{\mathrm{peak}}$
to visualize the achieved $\Delta$ for each case.
We find that increasing $\kappa$ and increasing $C$
have similar effects; they make both $\Delta$ and $\epsilon$ larger.
This is also confirmed in Fig.~\ref{fig:eps_and_delta},
where we plot the volume $\Delta$ (Top-Left)
and the lattice spacing $\epsilon$ (Top-Right)
against $C$ for various $\kappa$.
Note, however, that $\nu=\Delta / \epsilon$ behaves differently.
We find from Fig.~\ref{fig:eps_and_delta} (Bottom) that 
$\nu$ is almost independent of $C$,
while it increases for smaller $\kappa$.
Therefore, the basic strategy for tuning $\kappa$ and $C$
is to decrease $\kappa$ until 
$\nu$ reaches its maximum value\footnote{Note that the number
of data points of $R^{2}(t)$ is $(N-n+1)$ since the index
$K$ in (\ref{eq:def_t}) is restricted
as $0 \le K \le N-n$, where $n$ is the block size
used in defining (\ref{eq:def_abar}).
Therefore, $\nu$ cannot be larger than $(N-n+1)/2$.}
$(N-n+1)/2$
and to increase $C$ until an appropriate value of 
$\Delta$ or $\epsilon$ is achieved.
If we decrease $\kappa$ further, 
all the data points will lie in the region where the SSB of rotational 
symmetry occurs as in the situation with $\tilde{N}=16$
in Fig.~\ref{fig:test_rgm} (Right).
This is inconvenient since
we cannot determine the critical time $t_{\mathrm{c}}$ 
directly.


As practical applications of our method,
we can use it to investigate either the infinite volume 
limit ($\Delta\rightarrow \infty$ with fixed $\epsilon$) 
or the continuum 
limit ($\epsilon\rightarrow 0$ with fixed $\Delta$).
Figure \ref{fig:eps_and_delta} (Top-Left) and (Top-Right)
are useful for this purpose
since one can determine the value of $C$, which gives
the desired $\Delta$ and $\epsilon$ for a given $\kappa$.
As we mentioned earlier, 
$\nu$ can be as large as $(N-n+1)/2$ in the effective theory
by tuning $\kappa$, while it is 
$O(\sqrt{N})$ in the original model. 
This implies that the computational cost for the supersymmetric models
can be reduced from $O(N^5)$ to $O(N^{5/2})$
by the use of the effective theory with the optimal $\kappa$.


\section{Summary and discussions}
\label{sec:summary}

In this paper we have proposed a new method for probing
the dynamics of the Lorentzian type IIB matrix model
after the critical time, at which the spatial rotational symmetry
is spontaneously broken.
The basic idea is to consider the effective theory for the submatrices
corresponding to the time region after the critical time.
In particular, we make full use of the Lorentz invariance
to fix the form of the action for the effective theory.
This ansatz is verified by explicit calculations 
in the simplified models
despite the fact that the Lorentz invariance is softly broken by the 
infrared cutoffs, which are inevitably introduced to make the
models well defined.
We have also demonstrated the usefulness of the method
by reproducing the power-law behavior of the spatial extent 
observed in the bosonic model.
By tuning the parameters in the effective theory, one can 
obtain the information of the original supersymmetric model 
with the matrix size $N$ with the cost of $O(N^{5/2})$,
which is less than the cost for the bosonic models.
We therefore consider that this is a major achievement 
in the simulation method for the Lorentzian type IIB matrix model.
It is expected,
for instance, 
that the analysis performed in Ref.~\cite{Ito:2015mxa}
for the bosonic model has become feasible for the 
original supersymmetric model. 
We hope to report on the results in future publications.

\section*{Acknowledgements}

We thank S.~Iso and Y.~Kitazawa 
for valuable comments and discussions.
This research was supported by MEXT as
``Priority Issue on Post-K computer'' 
(Elucidation of the Fundamental Laws and Evolution of the Universe) 
and 
Joint Institute for Computational Fundamental Science (JICFuS).
Computations were carried out
using computational resources of the K computer 
provided by the RIKEN Advanced Institute for Computational Science 
through the HPCI System Research project (Project ID:hp160210).
The supercomputer FX10 at University of Tokyo was used
in developing our code for parallel computing,
and computational resources such as
KEKCC, NTUA het clusters and FX10 at Kyushu University
were used for calculations in the appendix \ref{sec:appendix_6dsusy}.
T.~A.\ and A.~T.\ were supported in part by Grant-in-Aid 
for Scientific Research (No.~17K05425 and 15K05046, respectively)
from Japan Society for the Promotion of Science.

\appendix

\section{Detailed information used in making the plots}
\label{sec:appendix_details}

In this appendix we present the detailed information used
in making the plots in Figs.~\ref{fig:10d-bosonic-original},
\ref{fig:10d-bosonic}, \ref{fig:c-dep_and_k-dep} 
and \ref{fig:eps_and_delta}.
The information includes:
the block size $n$
used in 
defining the extent of space $R^2(t)$ in (\ref{eq:def_rsq})
and the ``moment of inertia tensor'' (\ref{eq:def_tij});
the values obtained for the critical time $t_{{\rm c}}$ 
and the corresponding 
extent of space $R(t_{\rm c})$;
the number $\nu$ of data points of 
$R^{2}(t)$
contained within $t_{\mathrm{c}}<t\leq t_{\mathrm{peak}}$;
the volume $\Delta$ and the lattice spacing $\epsilon$
in the temporal direction defined by
(\ref{eq:def_delta}) and (\ref{eq:def_eps}), respectively.
Tables \ref{table:10dbosonic-original} and
\ref{table:10dbosonic-eff-theory}
show the information for Figs.~\ref{fig:10d-bosonic-original} and
\ref{fig:10d-bosonic}, respectively,
whereas Table \ref{table:6dbosonic-eff-theory}
shows the information for Figs.~\ref{fig:c-dep_and_k-dep} 
and \ref{fig:eps_and_delta}.

\begin{table}[H]
\begin{centering}
\begin{tabular}{|c||c|r@{\extracolsep{0pt}.}lr@{\extracolsep{0pt}.}l|cr@{\extracolsep{0pt}.}lr@{\extracolsep{0pt}.}l|}
\hline 
$N$ & $n$ & \multicolumn{2}{c}{$t_{\mathrm{c}}$} & \multicolumn{2}{c|}{$R\left(t_{\mathrm{c}}\right)$} & $\nu$ & \multicolumn{2}{c}{$\Delta$} & \multicolumn{2}{c|}{$\epsilon$}\tabularnewline
\hline 
\hline 
256 & 24 & -0&82166(6) & 0&30045(3) & 33 & 2&7380(7) & 0&08297(2)\tabularnewline
384 & 28 & -0&76930(7) & 0&26580(3) & 37 & 2&8943(6) & 0&07823(2)\tabularnewline
512 & 32 & -0&76559(7) & 0&24578(3) & 42 & 3&1150(7) & 0&07417(2)\tabularnewline
\hline 
\end{tabular}
\par\end{centering}
\begin{centering}
\caption{The information related to Fig.~\ref{fig:10d-bosonic-original}
is given for each $N$.
\label{table:10dbosonic-original}
}
\par\end{centering}
\end{table}

\begin{table}[H]
\begin{centering}
\begin{tabular}{|cr@{\extracolsep{0pt}.}lc||c|r@{\extracolsep{0pt}.}lr@{\extracolsep{0pt}.}l|cr@{\extracolsep{0pt}.}lr@{\extracolsep{0pt}.}l|}
\hline 
$N$ & \multicolumn{2}{c}{$\kappa$} & $C$ & $n$ & \multicolumn{2}{c}{$t_{\mathrm{c}}$} & \multicolumn{2}{c|}{$R\left(t_{\mathrm{c}}\right)$} & $\nu$ & \multicolumn{2}{c}{$\Delta$} & \multicolumn{2}{c|}{$\epsilon$}\tabularnewline
\hline 
\hline 
256 & 1&0 & 50 & 32 & -0&76795(4) & 0&29690(9) & 60 & 2&5867(2) & 0&043111(3)\tabularnewline
256 & 2&0 & 50 & 26 & -0&75087(5) & 0&26948(7) & 44 & 2&8490(3) & 0&064750(6)\tabularnewline
256 & 1&0 & 100 & 32 & -0&71405(5) & 0&25310(13) & 55 & 2&7730(3) & 0&050418(5)\tabularnewline
\hline 
\end{tabular}
\par\end{centering}
\begin{centering}
\caption{The information related to Fig.~\ref{fig:10d-bosonic}
is given for each $N$, $\kappa$ and $C$. 
\label{table:10dbosonic-eff-theory}
}
\par\end{centering}
\end{table}

\begin{table}
\begin{centering}
\begin{tabular}{|cr@{\extracolsep{0pt}.}lc||c|r@{\extracolsep{0pt}.}lr@{\extracolsep{0pt}.}l|cr@{\extracolsep{0pt}.}lr@{\extracolsep{0pt}.}l|}
\hline 
$N$ & \multicolumn{2}{c}{$\kappa$} & $C$ & $n$ & \multicolumn{2}{c}{$t_{\mathrm{c}}$} & \multicolumn{2}{c|}{$R\left(t_{\mathrm{c}}\right)$} & $\nu$ & \multicolumn{2}{c}{$\Delta$} & \multicolumn{2}{c|}{$\epsilon$}\tabularnewline
\hline 
\hline 
32 & 1&0 & 0 & 12 & -0&8044(4) & 0&6559(8) & 8 & 1&225(1) & 0&1531(2)\tabularnewline
32 & 1&0 & 5 & 12 & -1&0577(3) & 0&5350(7) & 10 & 1&977(1) & 0&1976(1)\tabularnewline
32 & 1&0 & 10 & 10 & -0&9194(4) & 0&4491(9) & 9 & 2&049(2) & 0&2277(2)\tabularnewline
32 & 1&0 & 20 & 8 & -0&7922(7) & 0&3268(10) & 8 & 2&424(3) & 0&3030(4)\tabularnewline
32 & 1&0 & 30 & 8 & -0&7875(7) & 0&2238(6) & 8 & 3&515(5) & 0&4393(6)\tabularnewline
\hline 
32 & 2&0 & 0 & 10 & -1&1028(9) & 0&5618(7) & 8 & 1&963(2) & 0&2454(3)\tabularnewline
32 & 2&0 & 5 & 10 & -1&1114(8) & 0&4896(9) & 8 & 2&267(3) & 0&2833(4)\tabularnewline
32 & 2&0 & 10 & 8 & -0&9438(11) & 0&4085(13) & 7 & 2&311(4) & 0&3302(5)\tabularnewline
32 & 2&0 & 20 & 8 & -0&9443(11) & 0&3173(12) & 7 & 2&983(5) & 0&4261(7)\tabularnewline
32 & 2&0 & 30 & 8 & -0&9398(10) & 0&2197(7) & 7 & 4&272(7) & 0&6103(10)\tabularnewline
\hline 
32 & 3&0 & 0 & 8 & -0&9408(13) & 0&5296(12) & 6 & 1&774(3) & 0&2957(6)\tabularnewline
32 & 3&0 & 5 & 8 & -1&1365(11) & 0&4327(10) & 7 & 2&628(4) & 0&3754(6)\tabularnewline
32 & 3&0 & 10 & 8 & -1&1416(13) & 0&3860(9) & 7 & 2&953(5) & 0&4218(8)\tabularnewline
32 & 3&0 & 20 & 8 & -1&1449(13) & 0&2981(5) & 7 & 3&836(7) & 0&5480(9)\tabularnewline
32 & 3&0 & 30 & 6 & -0&9462(18) & 0&1857(5) & 6 & 5&082(14) & 0&8470(23)\tabularnewline
\hline 
\end{tabular} 
\par\end{centering}
\caption{The information related to Figs.~\ref{fig:c-dep_and_k-dep} 
and \ref{fig:eps_and_delta}
is given for each $N$, $\kappa$ and $C$.}
\label{table:6dbosonic-eff-theory}
\end{table}

\section{Results for the 6d supersymmetric model}
\label{sec:appendix_6dsusy}

In this appendix we show the results of the analysis
in section \ref{sec:eff-theory}
for the 6d version
of the supersymmetric model\footnote{See section 3 
of Ref.~\cite{Ito:2013ywa} for its precise definition.
In eqs.~(\ref{Z-Likkt3}) and (\ref{eq:effective_theory}),
the Pfaffian ${\rm Pf}\mathcal{M}$ should be replaced 
by the determinant ${\rm det}\mathcal{M}$, which is real 
but not positive semi-definite. 
In simulating (\ref{Z-Likkt3}) and (\ref{eq:effective_theory}),
we take the absolute value of ${\rm det}\mathcal{M}$ assuming
that configurations with negative ${\rm det}\mathcal{M}$
are negligible. See footnote \ref{footnote:phase-quench}.}.
This confirms, in particular,  
that our method works also in the case 
where fermionic matrices are included.

\begin{figure}[t]
\centering{}
\includegraphics[width=7.3cm]{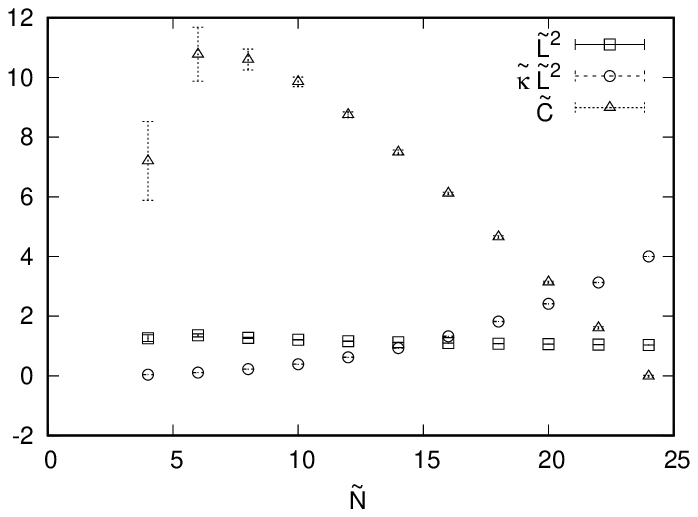}
\includegraphics[width=7.3cm]{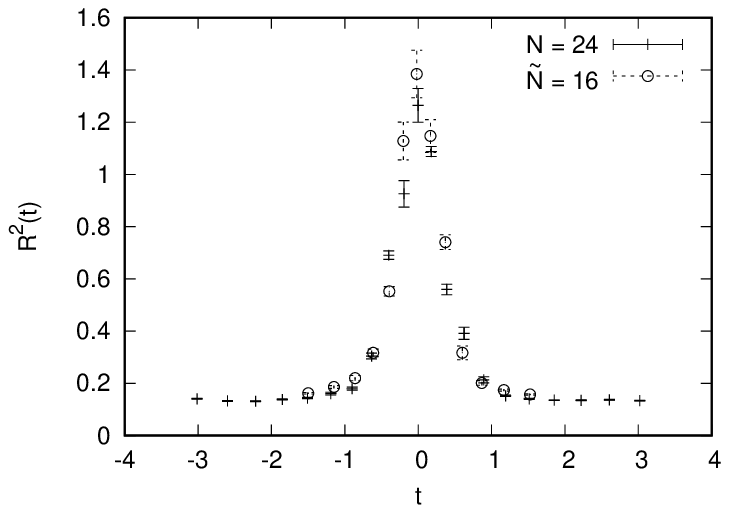}
\caption{
(Left) The expectation values $\tilde{L}^{2}$, $\tilde{\kappa}\tilde{L}^{2}$
and $\tilde{C}$ defined in (\ref{eq:observables}) are plotted against
the size $\tilde{N}$ of the submatrices
for the 6d supersymmetric model with $N=24$. 
(Right) The extent of space $R^{2}\left(t\right)$
is plotted against $t$ for the 6d supersymmetric 
model. The data points of plus sign represent 
the results for the original model with $N=24$. 
The circles represent the results of
$\tilde{R}^{2}\left(\tilde{t}\right)$ for the effective
theory with $\tilde{N}=16$.
The block size used in making this plot is $n=4$ for both cases. 
\label{fig:6d_susy}
}
\end{figure}

We perform simulation of this model with the matrix size $N=24$
and plot the expectation values 
(\ref{eq:observables}) for the $\tilde{N} \times \tilde{N}$ submatrices
against $\tilde{N}$ 
in Fig.~\ref{fig:6d_susy} (Left).
Then, using these expectation values,
we perform simulation of the effective theory
for the $\tilde{N} \times \tilde{N}$ submatrices.
In Fig.~\ref{fig:6d_susy} (Right),
we plot the extent of space $\tilde{R}^{2}(\tilde{t})$
against $\tilde{t}$ for the effective theory
with $\tilde{N}=16$ and 
compare it with the results for the original model with $N=24$.
We find that the effective theory reproduces the
late-time behaviors of the original model correctly.

\end{document}